\documentclass[12pt]{article}
\usepackage{amsmath,amssymb}
\newcommand{\bq}{\begin{eqnarray}}
\newcommand{\eq}{\end{eqnarray}}
\newcommand{\ov}{\overline}

\begin{document}

\begin{center}
{\large \bf A three-loop check of the "a - maximization"\\ in SQCD with adjoint(s)\,}
\end{center}

\begin{center}
{\bf Victor L. Chernyak}
\footnote
{\,\,\, E-mail: v.l.chernyak @ inp.nsk.su}
\end{center}

\begin{center}
{Budker Institute of Nuclear Physics,\,\, 630090, Lavrent'ev pr. 11, Novosibirsk, Russia}
\end{center}
\vspace{0.5cm}
\begin{center}{\bf Abstract}

\vspace{3mm}

The 'a - maximization" was introduced by K. Inrtiligator and B. Wecht for finding anomalous dimensions of chiral superfields
at the IR fixed points of the RG flow. Using known explicit calculations of anomalous dimensions in the perturbation theory
of SQCD (with one or two additional adjoint fields), it is checked here at the three-loop level.
\end{center}
\vspace{3mm}

The NSVZ $\beta$ - function in SUSY theory has the form \cite{NSVZ}\,:
\bq
\frac{d a}{d\ln \mu}=\beta(a)=-\frac{a^2}{1-a}\,N\,,\,\, N=\Biggl [ 3\, T_G-\sum_i T_{r_i}\Bigl (1+\gamma_i(a)  \Bigr )
\Biggr ], \,\, a=a_{NSVZ}=\frac{N_c g_{NSVZ}^2}{8\pi^2}\,,
\eq
where $\gamma_i$ are the field anomalous dimensions, $r_i$ denotes the representation of the i-th field. In what
follows we will consider two particular (but nontrivial) examples of the ${\cal N}=1 \,\,SU(N_c)$ gauge theory with
$N_F$ fundamentals $Q$ and $\ov Q$ and one ( $X$ ) or two ( $X,\,Y$ ) adjoint fields, and with zero superpotential.\\

\hspace*{1cm}{\bf One adjoint field} $X$\,. \quad Then\,:
\bq
N=\Biggl [b_o- N_F\gamma_Q(a)-N_c\gamma_X(a)\Biggr ]\,, \quad b_o=2N_c-N_F\,, \nonumber \\
\quad {\rm Tr}_r \Bigl (T^A T^B \Bigr )=T_r\,\delta^{AB}\,,\quad  T_G=T_X=N_c\,,\quad T_Q=T_{\ov Q}=\frac{1}{2}\,.
\eq
At the IR fixed point\,:
\bq
N=0,\quad  a=a^*=const,\quad d_i\equiv \Bigl (1-\frac{1}{2}\gamma_i(a^*)\Bigr )=\frac{3}{2}|R_i(a^*)|.
\eq

The problem is to find the values of all anomalous dimensions $\gamma_i(a^*)$ of chiral superfields, while $N=0$ gives
only one constraint. K. Intriligator and B. Wecht \cite{I1} found a method to find all $\gamma_i(a^*)$ (within the
conformal window).

As was shown in \cite{A1}, in SUSY theories at the conformal IR fixed point the well known central charge $a_T$ entering
the trace anomaly, $ T^{\mu}_{\mu}\sim a_T\,(\rm Euler)$\,, has the form\,:
\bq
a_T(R^*_i)=\sum_i \Biggl [ 3\Bigl (R^{(f)}_{i}(a^*) \Bigr )^3- R^{(f)}_{i}(a^*) \Biggr ],\quad R^*_i=R_i(a^*)\,,
\eq
where $R^{(f)}_{i}(a^*)$ is the superconformal R-charge of the fermionic component of the i-th superfield at the IR fixed
point. As was obtained in \cite{I1},  the right values $R_{i}^*$ maximize $a_T(R_i)$ (the method of
'a - maximization'\,, see also \cite{K1,I2,K2}\,). This allows one to find the values $R_i(a^*)$ and so $\gamma_i(a^*)$,
see (3).

The purpose of this short letter is to perform a check of the "a-maximization" predictions using the known results of explicit
calculations of $\gamma_i(a)$ up to a three-loop level in the perturbation theory of SQCD-like theories, see
\cite{J0,J1,J2} and refs. therein.
\footnote{\,
Our definition of $\gamma_i$ differs by the factor "-2" from those in \cite{J0}, i.e\,:
$\gamma_{\rm here}=-2\,\gamma_{\rm there}$\,.
}

The coupling is small in perturbation theory calculations. This is achieved at $z\equiv b_o/2N_c\ll 1$, i.e. at $N_F$
slightly below $2N_c$\,, and $N_c\gg 1$. So, the genuine small expansion parameter used everywhere below will be not
$a^*\ll 1$ by itself, but rather $z\ll 1$\,:
\bq
a^*(z)=a_1 z+a_2 z^2+a_3 z^3+O(z^4)\,.
\eq

We use below some notations from \cite{J0,J1,J2} (the dimensional reduction (DR) scheme is used therein)\,:
\bq
\Bigl (T^AT^A \Bigr )^i_j=C(r)^i_j=C_r\,\delta^i_j,\quad C_Q=C_{\ov Q}=C_F=\frac{N_c^2-1}{2N_c}\to \frac{N_c}{2},\quad
C_X=C_G=N_c\,,\\ \nonumber
T=\frac{1}{D}\sum_i {\rm Tr}_{r_i} \Bigl ( C(r_{i})^2 \Bigr ), \quad \Delta=\sum_{i}C_{r_i}\,T_{r_i}, \quad
D=N_c^2-1\,, \quad  \kappa=6\,\zeta(3).\\\nonumber
N_F=2N_c (1-z)\,,\quad 1/N_c\ll z=b_o/2N_c\ll 1\,,\quad  N_c\gg 1\,.
\eq

The perturbative series for the anomalous dimensions have the form\,:
\bq
\gamma^{NSVZ}_{i}(a)\equiv \gamma_i(a)=\gamma_i^{(1)}a+\gamma_i^{(2)}a^2+\gamma_i^{(3)}a^3\,,\\ \nonumber
\gamma^{DR}_i(a_{DR})\equiv \Gamma_i(a_{\rm DR})=\Gamma_i^{(1)}a_{\rm DR}+
\Gamma_i^{(2)}a_{\rm DR}^2+\Gamma_i^{(3)}a_{\rm DR}^3\,.
\eq

As shown in \cite{J2}, the couplings $a=a_{NSVZ}$ and $a_{DR}$ are connected (with the three-loop accuracy) as\,:
\bq
a_{\rm DR}=a\Biggl [1+a^2\Bigl ( -\frac{T}{2N_c^2}+\frac{b_o}{4N_c} \Bigr )  \Biggr ]\quad \to \quad
a^{*}_{\rm DR}=a^*\Biggl [1-a^{*2}\frac{T}{2N_c^2}  \Biggr ]+O(a^{*4})\,.
\eq

At the IR fixed point\,:
\bq
a(\mu/\Lambda\ll 1)\to a^*={\rm const},\quad  \Biggl [(1-z)\gamma_Q(a^*)+\frac{1}{2}\gamma_X(a^*)
\Biggr ]=z\,.
\eq

At one-loop\,:
\bq
\gamma_r^{(1)}=\Gamma_r^{(1)}=2\frac{C_r}{N_c}\,,\quad
\gamma_Q^{(1)}=2\frac{C_F}{N_c}\to 1,\quad \gamma_X^{(1)}=2\,,\quad \to \quad a^*(z)=\frac{z}{2}+O(z^2).
\eq

At two-loops\,:
\bq
\gamma_Q^{(2)}=\Gamma_Q^{(2)}= -\, \frac{1-2z}{2}\,,\quad \gamma_X^{(2)}=\Gamma_X^{(2)}= -\, 2(1-z)\,,\quad \to \quad
a^*(z)=\frac{z}{2}+ \frac{7}{16} z^2+O(z^3).
\eq

At three-loops \cite{J0,J1,J2}\,:
\bq
\Gamma_r^{(3)}=\kappa\,\frac{C_r}{N_c}\Biggl [-3+\frac{\Delta}{N_c^2}+\frac{b_o}{N_c} \Biggr ]+
\frac{C_r}{N_c}\Biggl [4\frac{C_r^2}{N_c^2}-5\frac{T}{N_c^2}-\frac{b_oC_r}{N_c^2}+\frac{5}{2}\frac{b_o}{N_c}
-\frac{b_o^2}{2N_c^2}  \Biggr ],\quad
\eq
\bq
\Gamma_Q^{(3)}\to \frac{\kappa}{2}\Biggl (-3+\frac{\Delta}{N_c^2}\Biggr ) +\frac{1}{2}\Biggl [ 1-5\frac{T}{N_c^2}\Biggr ],
\quad \Gamma_X^{(3)}\to\kappa \Biggl (-3+\frac{\Delta}{N_c^2} \Biggr )+ \Biggl [4-5\frac{T}{N_c^2} \Biggr ]\,,\quad
\eq
and from $\gamma(a^*)=\Gamma(a^{*}_{DR})$\,, see (8),(10)\,:
\bq
\gamma^{(3)}_r=\Gamma^{(3)}_r+\frac{C_r}{N_c}\Biggl (-\frac{T}{N_c^2}+\frac{b_o}{2N_c} \Biggr )\to
\Biggl [\Gamma^{(3)}_r-\frac{C_r}{N_c}\frac{T}{N_c^2}\Biggr ]\,,\nonumber \\
\gamma^{(3)}_Q \to \Biggl [\Gamma^{(3)}_Q-\frac{T}{2N_c^2}\Biggr ]\,,\quad
 \gamma^{(3)}_X \to \Biggl [\Gamma^{(3)}_X-\frac{T}{N_c^2}\Biggr ]\,.
\eq

It is convenient to define\,:
\bq
a^*={\ov a}+\frac{1}{4}\delta a^*\,,\quad {\ov a}=\frac{z}{2}\Bigl (1+\frac{7}{8}z  \Bigr )\,,\quad \delta a^*=O(z^3)\,.
\eq

Then one obtains from (9)\,:
\bq
\delta a^*=2z-\sum_{i=1}^{3}\,\Bigl ({\ov a} \Bigr )^i\Biggl [\gamma_X^{(i)}+2(1-z)\gamma_Q^{(i)}  \Biggr ]=
\frac{z^3}{8}\Biggl (\frac{15}{2}-\gamma_X^{(3)}-2\gamma_Q^{(3)}\Biggr )\,.
\eq

So\,:
\bq
\gamma_Q(a^*)=\Biggl [{\ov a}-(\frac{1}{2}-z)\,{\ov a}^2+\frac{15}{8}\,{\ov a}^3+\frac{{\ov a}^3}{2}
\Biggl (\gamma_Q^{(3)}-\frac{1}{2}\gamma_X^{(3)}\Biggr )  \Biggr ]\,, \nonumber \\
\gamma_X(a^*)=\Biggl [2\,{\ov a}-2\,(1-z)\,{\ov a}^2+\frac{15}{4}\,{\ov a}^3-{\ov a}^3
\Biggl (\gamma_Q^{(3)}-\frac{1}{2}\gamma_X^{(3)}\Biggr )  \Biggr ]\,,
\eq
and finally, see (13-15)\,:
\bq
\gamma_Q(a^*)=\Biggl [\frac{1}{2}z+\frac{5}{16}z^2+\frac{11}{64}z^3\Biggr ]+O(z^4)\,,\quad \gamma_X(a^*)=\Biggl [z+
\frac{3}{8}z^2+\frac{9}{32}z^3\Biggr ]+O(z^4)\,.\quad
\eq

Now, let us compare with the predictions of the 'a - maximization' \cite{I1}. These have the form
\cite{I1}\,\, ($\, x=N_c/N_F=1/(2-2z)$\,)\,:
\bq
 3R_Q(a^*)=\frac{x\Bigl (6x+3-\sqrt{20x^2-1}  \Bigr )-3}{2x^2-1}\,\,,\quad
3R_X(a^*)=\frac{10}{\Bigl (3+\sqrt{20x^2-1}  \Bigr )}\,\,.
\eq
Expanding these expressions in powers of $z$, one obtains\,:
\bq
3R_Q(a^*)=\Biggl [2-\frac{1}{2}z-\frac{5}{16}z^2-\frac{11}{64}z^3-\frac{109}{1024}z^4 \Biggr ],\,\,
3R_X(a^*)=\Biggl [2-z-\frac{3}{8}z^2-\frac{9}{32}z^3-\frac{67}{512}z^4 \Biggr ].\,
\eq

Finally, using\,: $\gamma_i(a^*)=\Bigl (2-3\,R_i(a^*) \Bigr )$\,, see (3), one finds that the 'a -maximization'
prediction agrees with the perturbation theory within the three-loop accuracy.

Attempts to use the results from \cite{J0,J1,J2} for a check of the 'a - maximization" predictions at two-  and three-loop
levels in SQCD with one adjoint have been made previously in \cite{K1,I2,K2}. They were successful only in part, i.e. only
the highest power terms $O(C_r^i)$ in $\gamma_r^{(i=2,3)}=A_r^{(i)}\,(C_r)^i+O(C_r^{i-1})$\, have been checked. The reason is
that the authors of \cite{K1,I2,K2} used expansions in powers of the coupling $a^*$\,, but $a^*$ does not enter
explicitly the answers (19) for $R_i$ ( and, besides, $a^*$ and $b_o/N_c$ are not independent).  To make a complete check, 
one has to use the genuine expansion parameter $z=b_o/2N_c \ll 1$.\\

\hspace*{1cm}{\bf Two adjoint fields} $X\,,\,Y$\,. \quad In this theory (the anomalous dimensions $\gamma_{X}=\gamma_{Y}$)\,:
\bq
b_o=N_c-N_F\,,\quad N=\Biggl [b_o-N_F\gamma^*_Q-2N_c\gamma^*_X\Biggr ]=0\,,\quad N_F=N_c(1-2z)\,,\quad \frac{1}{N_c}\ll
z\ll 1\,,
\eq
and instead of (9) one has now
\bq
\Bigl (\frac{1}{2}-z \Bigr )\gamma^*_Q+\gamma^*_X=z\,.
\eq

Proceeding in the same way as above, one obtains
\bq
a^*\equiv{\ov a}+\frac{2}{5}\delta a^*\,,\quad {\ov a}=\frac{2}{5}z\Bigl (1+\frac{19}{25}z \Bigr )\,,\quad
\delta a^*=\frac{4}{125}z^3 \Bigl (\frac{58}{5}-2\gamma_X^{(3)}-\gamma_Q^{(3)} \Bigr )\,,
\eq
where $\gamma_Q^{(3)}$ and $\gamma_X^{(3)}$ are given in (13,14). The anomalous dimensions $\gamma_Q^{(3)}$ and
$\gamma_X^{(3)}$ are obtained now as
\bq
\gamma_Q^*=\Biggl [\Bigl ({\ov a}+\frac{2}{5}\delta a^*\Bigr )+(z-\frac{1}{2}){\ov a}^2+\gamma_Q^{(3)}{\ov a}^3  \Biggr ]=
\Biggl [\frac{2}{5}z+\frac{28}{125}z^2+\frac{344}{3125}z^3 \Biggr ]+O(z^4),\nonumber
\eq
\bq
\gamma_X^*=\Biggl [2\Bigl ({\ov a}+\frac{2}{5}\delta a^*\Bigr )+(2z-2){\ov a}^2+\gamma_X^{(3)}{\ov a}^3 \Biggr ]=
\Biggl [\frac{4}{5}z+\frac{36}{125}z^2+\frac{528}{3125}z^3 \Biggr ]+O(z^4).
\eq

The "a-maximization" \cite{I1} predicts in this case the values of $R_i^*$-charges as:
\bq
3R_Q^*=\Biggl [\, 3+\frac{3x-2x\sqrt{26x^2-1}}{(8x^2-1)}\,\Biggr ],\quad 3R_X^*=\frac{1}{2}\Biggl [\, 3-\frac{3-2\sqrt
{26x^2-1}}{(8x^2-1)}\,  \Biggr ]\,,
\eq
with $x=N_c/N_F=1/(1-2z)$. Finally, expanding (25) in powers of z, one obtains:
\bq
3R_Q^*=\Biggl [2- \frac{2}{5}z-\frac{28}{125}z^2-\frac{344}{3125}z^3-\frac{928}{15625}z^4 \Biggr ],\quad
3R_X^*=\Biggl [2-\frac{4}{5}z-\frac{36}{125}z^2-\frac{528}{3125}z^3-\frac{1256}{15625}z^4\Biggr ].
\eq
So, because $\gamma^*_i=(2-3R^*_i)$\,, this agrees with (24), within the three-loop accuracy.\\

\vspace{2mm}

This work is supported in part by the RFBR grant 07-02-00361-a.\\

\end{document}